\documentclass[12pt]{article}

\usepackage{graphicx}
\graphicspath{ {./images/} }

\usepackage{amssymb,amsmath}
\usepackage{cite}

\begin{document}

\begin{center}
\phantom{0}
\vskip -20mm

\vskip 10mm

{\bf
Alexander Andrianov, Yasser Elmahalawy and Artem Starodubtsev
\footnote{St.Petersburg state university, St.Petersburg, Russia, email: sashaandrianov@gmail.com, yasserreda99@gmail.com, artemstarodubtsev@gmail.com}}
\vskip 0.5em

{\bf
 2+1-dimensional gravity coupled to a dust shell: quantization in terms of global phase space variables.
}
\end{center}

\begin{abstract}
We perform canonical analysis of a model in which gravity is coupled to a spherically symmetric dust shell in 2+1 spacetime dimensions. The result is a reduced action depending on a finite number of degrees of freedom. The emphasis is made on finding canonical variables providing the global chart for the entire phase space of the model. It turns out that all the distinct pieces of momentum space could be assembled into a single manifold which has $ADS^2$-geometry, and the global chart for it is provided by the Euler angles.   This results in both non-commutativity and discreteness in coordinate space, which allows to resolve the central singularity. We also find the map between  $ADS^2$ momentum space obtained here and momentum space in Kuchar variables, which could be helpful in extending the present results to 3+1 dimensions.
\end{abstract}

\section{Introduction}
It has long been believed that quantum theory of gravity could regularize its singularities. The first argument for that dates back to Bronstein \cite{mpb}, who showed that there could be no measurable distances smaller than the Planck length.
In the absence of a full theory of gravity there is still no decisive answer whether this is indeed the case or not.

One can address this question by studying reduced models for General Relativity in which all but a few degrees of freedom are removed. Of special interest are those which contain black hole  solutions, which are essential for Bronstein's argument.

The simplest model possible is a homogenous universe with a matter field. Quantization of this model has been extensively studied, and the overall conclusion is that quantum theory does't cure the singularity in this case \cite{lqc}, unless we include some exotic matter\cite{an}.

The second simplest model is spherically symmetric spacetime in which matter is represented by one or more dust shells.
Unlike homogenous  model, it accomodates black hole like solutions, which results in a non-trivial phase space with branching of the solution to the constraints.

There is a variety of works studying such models both on classical \cite{israel1,kuchar,hajicek } and quantum \cite{louko,vb,hk} level.
In some of the versions of quantum theory the central singularity is removed \cite{vb,hk}. However  the above results
do not always agree with each other. Apart from quantization ambiguity, the other possible reason for that is a complicated phase space structure of the model. Different quantum theories could arise on different sectors of such phase space.

In such situation the common wisdom is that the wavefunction of a quantum theory has to be defined on all possible configurations, independently of whether they are classically accessible or not.
In a particular way it was realized in \cite{vb} where by making use of complex  coordinates different sectors of the phase space were assembled into one Riemann surface, the branching point identified with a horizon.

Another possibility is to try to find a real global chart for the phase space (if it exists). One example where it was possible to realize is 2+1 dimensional gravity coupled to a point particle \cite{thooft}. The momentum of the particle turn
out to live on the Lorentz group manifold, and the different branches of the solution to the constraints result from different way of intersecting this manifold by a plane.

This will be our starting point in attempting to relate the two above approaches.

In section 2 we repeat the canonical analysis \cite{vb,hajicek}  for gravity+shell system, including Kuchar canonical transformation necessary for it, but now in 2+1 spacetime dimensions. The result is very similar to that in 3+1 dimensions, the only difference being that there is no Newtonian potential, but the branching of the constraint solution is the same.

In section 3 we extend the results of  \cite{thooft} and \cite{mw} from a point particle to a spherical shell, representing the later as an ensemble of infinite number of point particles. The momentum space of the shell turn out to be
$ADS^2$, which results, in particular, in non-commutativity of the coordinates. The Hamiltonian constraint is, however, different from that of a particle, which accounts for a contribution of inter-shell movement energy to the gravitational field.
The relation between momenta from $ADS^2$ and canonical momenta from \cite{vb,hajicek} is found.

In section 4 we perform quantization in momentum  representation on $ADS^2$. Apart from non-commutativity of the coordinates it results in discreteness of the spectrum of one of them (time). The areal radius of the shell has discrete spectrum when it is timelike and continuous, but separated from zero spectrum when it is spacelike. The later provides a mechanism for resolution of the central singularity.

Finally, the implementation of the Hamiltonian constraint and finding the physical Hilbert space is discussed.
We also discuss a possibility to extend the above results to 3+1 spacetime dimensions.

\section{Canonical analysis of gravity+shell action and the problem of finding the global phase space variables}\label{secondorder}
In this section we rederive the results of \cite{vb,hajicek} in space one dimension lower.

\subsection{Canonical Formalism for Spherically Symmetric Spacetimes in 2+1 Gravity}
The ADM apporach to (2+1)-dimensional starts with a slicing of the spacetime manifold $M$ into constant-time surfaces $\Sigma$ and an induced metric $g_{ij}$. Consider a spherically symmetric two-dimensional Riemannian space $(\Sigma,g)$. The line element $d\sigma$ on $\Sigma$ is characterized by two functions $\Lambda(r)$ and $R(r)$,
\begin{align}
\label{1}
d\sigma^{2} & =   \Lambda^{2}(r) dr^{2} + R^{2}(r) d\theta^{2}.
\end{align}
We take $\Lambda(r)$ and $R(r)$ to be positive and non-vanishing functions. Note that $R(r)$ is the curvature radius and $ d\sigma = \Lambda(r) dr$ is the radial line element. The Dirac-ADM action for vacuum metric is given by
\begin{align}
\label{2}
S _{\Sigma}[g,N,N^{a}] & =   \frac{1}{16\pi } \int _{\mathcal{M}} R \sqrt{-g} d^{3}x = \int dt  \int _{\Sigma} L _{\Sigma} d^{2}x,
\end{align}
which represent the standard Einstein-Hilbert action for the gravitational field and Lagrangian $L _{\Sigma}$ is
\begin{align}
\label{3}
L _{\Sigma} & =   \frac{N}{16\pi } (K^{ab}K_{ab} - K^2 + \bold{R}[g]),
\end{align}
where $\bold{R}[g]$ is the scalar curvature of a space metric $g_{ab}=(\Lambda^2, R^2)$ and $K_{ab}$ is the extrinsic curvature for constant time surface. The scalar curvature for line element \ref{1} is
\begin{align}
\label{4}
\bold{R}[g] & =   -2 \Lambda^{-2} R^{-1}R'' + 2 \Lambda^{-3}R^{-1} \Lambda' R'.
\end{align}
Note that dot and prime denote differentiation with respect to t and r, respectively. The extrinsic curvature is defined by
\begin{align}
\label{4a}
K_{ij} & =   \frac{1}{2N}  \Big( - \partial_{t}g_{ij}  + {}^{(2)} \nabla_{i} N_{j} + {}^{(2)} \nabla_{j} N_{i} \Big),
\end{align}
where $ {}^{(2)} \nabla$ is the full covariant derivative and shift vector $N^i = (N^r,0)$. For spherically symmetric, the extrinsic curvature for line element \ref{1} is diagonal:
\begin{align}
\label{5}
K_{i}^{j} & =  diag(K_{r}^{r}, K_{\theta}^{\theta}  ), \\
\label{5a}
K_{rr} & =   -N^{-1} \Lambda \Big(\dot{\Lambda} - (\Lambda N^{r})'\Big), \\
\label{6}
K_{\theta\theta} & =   -N^{-1} R (\dot{R} - R' N^{r}).
\end{align}
The ADM action in terms of Eqs. \ref{3}, \ref{4}, \ref{5a} and \ref{6} is
\begin{align}
\label{7}
S _{\Sigma}[R,\Lambda;N,N^{a}] & = \int dt  \int _{-\infty}^{\infty} dr \Bigg[ -N^{-1}  \Big(-\dot{\Lambda} +  (\Lambda N^{r})'\Big) \Big(\dot{R} +R' N^{r}\Big) \nonumber  \\ &+  N \Big(- \Lambda^{-1} R'' +  \Lambda^{-2} R' \Lambda '\Big)  \Bigg].
\end{align}
The canonical formalism of the action is derived by differentiating the ADM action \ref{7} with respect to velocities $\dot{\Lambda}$ and $\dot{R}$, we obtain
\begin{align}
\label{8}
P_{\Lambda} & =  -N^{-1} (\dot{R} - R' N^{r}), \\
\label{9}
P_{R} & =   -N^{-1} \Big(\dot{\Lambda} - (\Lambda N^{r})'\Big),
\end{align}
where the momentum $P_{R}$ is a density, while the momentum $P_{\Lambda}$ is a scalar. The velocities $\dot{\Lambda}$ and $\dot{R}$ can be written in terms of $P_{R}$ and  $P_{\Lambda}$ as
\begin{align}
\label{10}
\dot{R} & =  -N P_{\Lambda}  + R' N^{r}, \\
\label{11}
\dot{\Lambda} & =   -N P_{R} + (\Lambda N^{r})'.
\end{align}
The extrinsic curvature as a function of the canonical momenta:
\begin{align}
\label{11a}
K_{rr} & =   \Lambda P_{R}, ~~~~~  K_{\theta\theta}  =  R P_{\Lambda}.
\end{align}
The ADM action \ref{7} can be written in the canonical form by Legendre transformation as
\begin{align}
\label{12}
S _{\Sigma}[R,\Lambda,P_{\Lambda},P_{R};N,N^{r}] & = \int dt  \int _{-\infty}^{\infty} dr (P_{\Lambda} \dot{\Lambda} + P_{R} \dot{R} - NH^{G} - N^{r}H_{r}^{G} ),
\end{align}
where  the super-Hamiltonian is
\begin{align}
\label{12a}
H^{G} & = - P_{\Lambda}P_{R} + \Lambda^{-1} R'' -  \Lambda^{-2} R' \Lambda',
\end{align}
and supermomentum
\begin{align}
\label{12b}
H_{r}^{G} & = P_{R}R' - \Lambda P'_{\Lambda}.
\end{align}

\subsection{Kucha\v r Transformation for 2+1 Gravity}
In \cite{kuchar}, Kucha\v r proposed a new method to express Hamiltonian and momentum constraints into simple set of constraints in (3+1)-dimensional gravity. This method is based on canonical transformation of the old variables to new canonical set. In this section, we discuss  Kucha\v r method in (2+1)-dimensional gravity. The general metric of a spherically symmetric spacetime \cite{hajicek,vb} is
\begin{align}
\label{13}
ds^{2} & = -\big( N^{2} - \Lambda^{2} (N^{r})^{2} \big) dt^{2} + 2 \Lambda^{2} N^{r} dtdr + \Lambda^{2}dr^{2} + R^{2} d \theta^{2},
\end{align}
where $N, N^{r}, \Lambda$ and $R$ are continuous fuctions of $t$ and $r$ only. Eq. \ref{13} represent the Arnowitt-Deser-Misner (ADM) form of the 2+1 decomposition of a space-time. The general spherically symmetric (2+1)-dimensional Schwarzchild metric, in the curvature coordinates $(T,R)$, is
\begin{align}
\label{14}
ds^{2} & =  -(1-2m) dT^{2} + (1-2m)^{-1}dR^{2} +R^{2} d \theta^{2},
\end{align}
where conical singularity with angle $\sim 1- \sqrt{1 - 2m} \simeq M$. Consider that the hypersurface be a leaf of a foliation
\begin{align}
\label{14a}
T & = T(t,r), ~~~~~ R=R(t,r).
\end{align}
We substitute Eq. \ref{14a} into  Eq. \ref{14} and get
\begin{align}
\label{14b}
ds^{2} & =  - \Big[ (1-2m) \dot{T}^{2} - (1-2m)^{-1}\dot{R}^{2}  \Big] dt^{2} \nonumber  \\ &+2 \Big[ - (1-2m) \dot{T} T' + (1-2m)^{-1}\dot{R} R'  \Big] dtdr \nonumber  \\ &+  \Big[ -(1-2m) T^{'2} + (1-2m)^{-1} R^{'2}  \Big] dr^{2} +R^{2} d \theta^{2}.
\end{align}
By comparing  Eq. \ref{14b} with  Eq. \ref{13}, we get
\begin{align}
\label{14c}
 \Lambda^{2} & =-(1-2m) T^{'2} + (1-2m)^{-1} R^{'2},   \\
\label{14d}
\Lambda^{2} N^{r} & = - (1-2m) \dot{T} T' + (1-2m)^{-1}\dot{R} R',  \\
\label{14e}
N^{2} - \Lambda^{2} (N^{r})^{2} & =(1-2m) \dot{T}^{2} - (1-2m)^{-1}\dot{R}^{2}.
\end{align}
The lapse function and shift function are given by
\begin{align}
\label{14f}
N^{r} & = \frac{- (1-2m) \dot{T} T' + (1-2m)^{-1}\dot{R} R'}{-(1-2m) T^{'2} + (1-2m)^{-1} R^{'2}},   \\
\label{14g}
 N & = \frac{ \dot{T} R' - \dot{R} T'}{\sqrt{-(1-2m) T^{'2} + (1-2m)^{-1} R^{'2}}}
\end{align}
,respectively. To calculate $P_{\Lambda}$, we substitute Eqs. \ref{14f} and \ref{14g} into Eq. \ref{8}
\begin{align}
\label{14h}
-T' & = (1-2m)^{-1} \Lambda P_{\Lambda}.
\end{align}
The Schwarzchild mass can be calculated using Eqs. \ref{14c}, which is equal to
\begin{align}
\label{14i}
m& = \frac{1}{2} + \frac{P_{\Lambda}^{2}}{2} - \frac{1}{2}\frac{R^{'2}}{\Lambda^{2}}.
\end{align}
We found that functions $m(r)$ and $-T'(r)$ are canonically conjugate variables and the dynamical variable $-T'(r)$ can be denoted by $P_{m}(r)$. The new momentum $\bar{P}_{R}(r)$ is written interms of the old momentum $P_{R}(r)$ and a dynamical variable $\Phi(r)$,
\begin{align}
\label{14j}
\bar{P}_{R}(r) & = P_{R}(r) +\Phi(r;R,\Lambda,P_{\Lambda}],
\end{align}
where $\Phi(r)$ does not depend on $P_{R}(r)$. Finally, the transformation form is
\begin{align}
\label{15a}
\Lambda & =  \sqrt{ R^{'2} (1-2m)^{-1}- (1-2m) P^2_m},   \\
\label{15b}
P_{\Lambda} & =\frac{ (1-2m)P_{m}}{\sqrt{ R^{'2} (1-2m)^{-1} - (1-2m) P^2_m}},   \\
\label{15c}
\bar{R} & =  R,  \\
\label{15d}
\bar{P}_{R} & =  P_R -  \frac{(1-2m)^{-1}}{\Lambda^{2}}   \Big[ (\Lambda P_{\Lambda})' R' - (\Lambda P_{\Lambda}) R''  \Big].
\end{align}
The  transformation form is not valid at the Horizon. We used natural units in which $G=c=1$. The Liouville form is
\begin{align}
\label{16}
\Theta & =  \int   P_{\Lambda}  \dot{\Lambda} + P_{R}  \dot{R}  \\
\label{17}
 & =  \int   P_{m}  \dot{m} + \bar{P}_{R}  \dot{R} + \frac{\partial}{\partial t} \Bigg[ \Lambda P_{\Lambda} + \frac{R'}{2} \ln  \Bigg| \frac{R' - \Lambda P_{\Lambda}}{R' + \Lambda P_{\Lambda}}  \Bigg| \Bigg] \nonumber  \\ & +  \frac{\partial}{\partial r} \Bigg[ \frac{ \dot{R}}{2} \ln \Bigg| \frac{R' + \Lambda P_{\Lambda}}{R' - \Lambda P_{\Lambda}}  \Bigg| \Bigg].
\end{align}
Eqs, \ref{12a} and \ref{12b} can be represented by simple set of constraints:
\begin{align}
\label{18}
\bar{P}_{R} & =0, ~~~~~ m'=o.
\end{align}
except on the horizon.

\subsection{Canonical analysis for (2+1)-dimensional gravity with thin shell}
In this section, we briefly describe the canonial formalism for a spherically symmetric gravitational field of the 2+1 decomposition of a space-time with thin shell. The action for this system is
\begin{align}
\label{19}
S & =  S_{gr} + boundary~ terms + S_{shell}   \nonumber  \\ & = \frac{1}{16\pi G} \int  R \sqrt{-g} d^{4}x + (boundary~ terms) + M \int _{\Sigma}  d\tau.
\end{align}
The first and third term in Eq. \ref{19} represent the standard Einstein-Hilbert action for the gravitational field and dust thin shell action, respectively. The standard Einstein-Hilbert action is given by Eq. \ref{19} and shell part of the action is
\begin{align}
\label{20a}
S_{shell} & =  -  m \int _{\Sigma} \sqrt{\hat{N}^{2} - \hat{\Lambda}^{2}(\hat{N}^{r}+\dot{\hat{r}})^{2}}  dt,
\end{align}
where hats denote the value of variables on the shell and $m$ is the rest mass of the shell. The explicit form of the action \ref{19} in the Hamiltonian form becomes
\begin{align}
\label{20}
S & = \int dt  \hat{\pi} \dot{\hat{r}} +  \int  \Big[ P_{\Lambda} \dot{\Lambda} + P_{R} \dot{R} - N (H^{s} +H ^{G}) - N^{r} (H_{r}^{s} - H_{r}^{G}) \Big] dr dt + \int dt M_{ADM},
\end{align}
where $M_{ADM}$ is the total mass of the combined gravity-shell system and $ \hat{\pi}$ is the momentum conjugate to $ \dot{\hat{r}}$ which is equal to
\begin{align}
\label{21}
\hat{\pi} & = \frac{m \hat{\Lambda}^{2} (\hat{N}^{r}+\dot{\hat{r}}) }{\sqrt{\hat{N}^{2} - \hat{\Lambda}^{2}(\hat{N}^{r}+\dot{\hat{r}})^{2}}},
\end{align}
and the super-Hamiltonian of the shell is
\begin{align}
\label{22}
H^{s} & = \sqrt{(\hat{\pi}/\hat{\Lambda})^{2} + m^{2}}\delta(r- \hat{r}),
\end{align}
and supermomentum of the shell is
\begin{align}
\label{23}
H_{r}^{s} & = \hat{\pi} \delta(r- \hat{r}).
\end{align}
The regular contribution to the constraints is the same as in vacuum and is valid in inner and outer regions of the shell.
There is also a delta-functional contribution on the shell which has to be combined with the shell hamiltonian. As a result we get the shell constraints:
\begin{align}
\label{24}
C^{s} & = \frac{[R']}{\Lambda}+\sqrt{(\hat{\pi}/\hat{\Lambda})^{2} + M^{2}},
\end{align}
and
\begin{align}
\label{25}
C_{r}^{s} & =\Lambda [P_\Lambda] +\hat{\pi},
\end{align}
where the square brackets mean the jump of the field across the shell.

Now we go to the Kuchar variables, solve the constraints in the inner and outer regions, and plug the solution back into the action. It turns out that the the kinetic term in the bulk disappears due to the bulk constraints, kinetic term   containing  $ \dot{\hat{r}}$ on the shell disappears due to the constraint (\ref{25}), and all that remains is the contribution from boundary terms that appear after the Kuchar canonical transformation.
\begin{align}
\label{26}
S & =   \int dt \Big[ m \dot {\hat T} + [P_{R}] \dot{R} -N^s C^s \Big],
\end{align}
Here $ {\hat T}$ is the Killing time evaluated at the shell,
\begin{align}
\label{27}
{P_R}\Big\vert_{in,out}=\ln \Bigg| \frac{R' + \Lambda P_{\Lambda}}{R' - \Lambda P_{\Lambda}}  \Bigg|_{in,out},
\end{align}
and $C^s$ is the constraint (\ref{25}).

Then we express the  constraint (\ref{25}) in terms of the shell canonical variables $m$ and $P_R$:
\begin{align}
\label{28}
C^s=\sqrt{1-2m}\cosh{P_R}_{out} -\cosh{P_R}_{in}+M
\end{align}
From the resulting action we find equations of motion for $R$
\begin{align}
\label{29}
\frac{\dot R}{N^s}=\sqrt{1-2m}\sinh{P_R}_{out}=\sinh{P_R}_{in},
\end{align}
which leads to another constraint
\begin{align}
\label{30}
\sqrt{1-2m}\sinh{P_R}_{out}-\sinh{P_R}_{in}=0.
\end{align}
Unlike 3+1 dimensional situation the constraints (\ref{28}) and (\ref{30}) are now first class.
Substituting (\ref{29}) into (\ref{28}) we recover the Israel equation for the shell:
\begin{equation}\label{31}
\sqrt{1+\frac{\dot R^2}{(N^s)^2}}+\sqrt{1-2m+\frac{\dot R^2}{(N^s)^2}}-M=0.
\end{equation}

Finally, squaring the two constraints (\ref{28}) and (\ref{30}) and adding them we find the single Hamiltonian constraint describing the dynamics of the shell:
\begin{align}
\label{32}
1+1-2m-2\sqrt{1-2m}\cosh [P_R]-M^2=0.
\end{align}
This is the constraint to be used in quantum theory.

Here we have to point out that many of the above equations contain square roots in them which are not single-valued functions. Different choices of the signs in front of square roots correspond to different sectors of the phase space of the model, which are pictured as different regions on Penrose diagrams. The same $m$ corresponds to two different points in momentum space. Besides the Killing time $\hat T$ and the radial momentum $[P_R]$ diverge at the gravitational collapse point (horizon).
Beyond this point equation (\ref{32}) does not have solutions in real variables.
In other words the theory is formulated in variables which do not cover the phase space globally.

One way to circumvent the problem is to introduce a complex phase space ant to assemble different patches into a Riemann surface \cite{vb}.

However there is still a possibility that there exists another set of real phase space variables which would cover it globally. This possibility is studied in the next section.

\section{First order formalism and anti-DeSitter the momentum space}\label{firstorder}
There is an example in literature  when a simple gravity+matter model was given a global chart of its phase space.
This is 2+1-dimensional gravity coupled to point particles \cite{thooft}. In this section we generalize these results to a spherically symmetric shell.

\subsection{Action principle and phase space reduction}
In this section we start with a generic (non-spherically symmetric) action for 2+1 gravity coupled to a finite number of particles.

The basic variables are the triad $e_\mu=e_\mu^a\gamma_a$ and the connection $\omega_\mu^{ab}\gamma_a\gamma_b$, where $\gamma^a$ are generators of sl(2)-algebra.
The action reads:
\begin{equation}\label{action3d}
S=\int_M d^3 x \epsilon^{\mu \nu \rho} Tr (e_\mu R_{\nu \rho}) + S_{shell}
\end{equation}
where $R_{\nu \rho}$ is the curvature of $\omega_\rho$.

The shell will be discretized ( Consisting of $N$ particles labelled by index $i$)
\begin{equation}\label{actionshell}
S_{shell}=\sum\limits_i^N \int_{l_i} Tr(K_i e_\mu) dx^\mu
\end{equation}
where $l_i$ is i-th particle worldline and $K_i=m_i \gamma_0$ -- a fixed element of sl(2)-algebra.
The condition of spherical symmetry will be imposed later.

Gravity action is invariant with respect to gauge transformations:

\begin{equation}\label{gaugetrans}
\omega_\mu \rightarrow g^{-1}(\partial_\mu+\omega_\mu)g \ \ \ \ \
e_\mu \rightarrow g^{-1}(e_\mu + \partial_\mu \xi)g
\end{equation}
where $g$ is an SL(2) group element, and $\xi$ is an  sl(2) algebra element.

The shell action is not invariant.  The i-th particle term transforms as
\begin{equation}\label{wbg}
\int_{l_i} Tr(K_i e_\mu)dx^\mu \rightarrow \int_{l_i} Tr(\tilde K_i e_\mu)dx^\mu+\int_{l_i} Tr(\tilde K_i \dot \xi)d\tau
\end{equation}
where $\tilde K_i= gK_ig^{-1}$, $\tau$ is a parameter along the particle worldline and dot is the derivative with respect to it.

In the last term in the r.h.s. of \ref{wbg} one can recognize the standard particle action as it has the form of
$\int p_a\dot x^a$, where $p_a=Tr (\gamma_a \tilde K_i)$, $x^a=Tr(\gamma^a \xi)$, and given the definition of $\tilde K_i$
$p^a$ satisfies the constraint $p^a p_a=m^2$. Thus the particles degrees of freedom are represented as gauge degrees of freedom evaluated at the location of the particles.

As before, to obtain a reduced action for this model  we have to solve the constraints and plug the solution back into the initial action. We choose slicing so that particle worldlines move along the time coordinate and obtain the constraints
by varying action (\ref{action3d}) with respect to $\omega_0$ and $e_0$:

\begin{equation}\label{const1st}
\epsilon^{0\mu \nu}\nabla_\mu e_\nu=0 \ \ \ \ \ \epsilon^{0\mu \nu}R_{\mu \nu}=\sum\limits_i^N \tilde K_i \delta^2 (x,x_i)
\end{equation}
where $x_i$ is the location of the i-th particle.
The first constraint (\ref{const1st}) generates the first of the transformations (\ref{gaugetrans}) and the second generates the second.

By using transformations (\ref{gaugetrans}) one can put to zero simultaniously one component of $\omega$ and one component of
$e$. This automatically linearizes the constraints  (\ref{const1st}). However, such  a gauge choice cannot be made globally,
because the model has a non-trivial moduli space, containing for example the gauge parameter evaluated at the location of one particle with respect to another. Following \cite{amms} we divide the  spacial slice into regions in each of which the above gauge choice could be made. Each such region should contain no more than one particle. Around each particle we draw a circle, so that  the circle are  connected to a common origin, but have no common boundaries, as it is shown on Fig.1. By making cuts along the circles the manifold is divided into $N$ discs, each containing a particle, and a polygon containing no particles, but connected to infinity.
\begin{figure}
\includegraphics[scale=0.70]{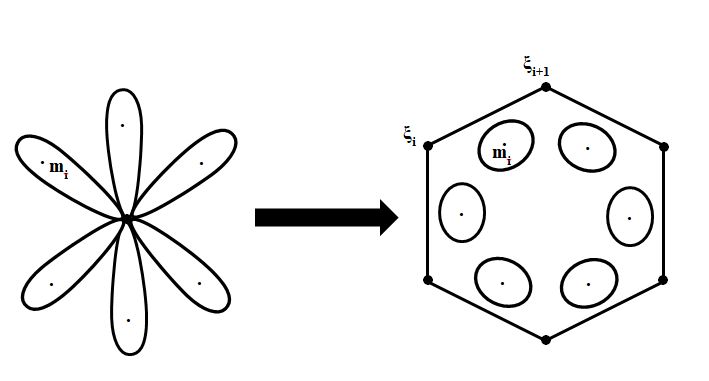}
\centering
\caption{Dividing space into discs and a polygon}
\end{figure}
For the discs it is convenient to write down the solution in polar coordinates with the origin at the location of the particles. We choose the gauge in which the radial components of $e$ and $\omega$ equal zero, solve the constraints,
and put the solution back into an arbitrary gauge:
\begin{equation}\nonumber
\omega_{r,i}=g_i^{-1}\partial_r g_i \ \ \ \ \omega_{\phi,i}=g_i^{-1}\nabla_\phi g_i
\end{equation}
\begin{equation}\label{constsol}
e_{r,i}=g_i^{-1}\partial_r \xi_i g_i \ \ \ \ e_{\phi,i}=g_i^{-1}\nabla_\phi \xi_i g_i
\end{equation}

where $\nabla_\phi \xi_i=\partial_\phi \xi_i+[\xi_i,K_i ]$.

And similar for polygon, for which the gauge parameters will be denoted $h$ and $\zeta$.

Now this solutions have to be put back into the kinetic term of the action which reads (for i-th disk):
\begin{equation}\label{kinterm}
S_{D_i}=\int_{D_i} d^3x \epsilon^{0\mu \nu} Tr(e_\mu \dot \omega_\nu) + \int_{l_i} Tr(\tilde K_i \dot \xi_i)d\tau
\end{equation}
By using the identity (notice that $K_i$ does not depend on time, so $\nabla_\phi$ commutes with time derivative)
\begin{equation}\nonumber
\dot{g_i^{-1}\nabla_\mu g_i}=g_i^{-1}\nabla_\mu(\dot g_i g_i^{-1}) g_i
\end{equation}
we find that in the first term of (\ref{kinterm}) there is a $\delta$-functional contribution which cancels the second
term, plus another term which is a total derivative. Thus the action for the disk collapses to its boundary:
\begin{equation}\label{kintermb}
S_{ D_i}=\int_{\partial D_i} d^2x  Tr(\nabla_\phi \xi_i \dot g_i g_i^{-1})
\end{equation}
Similar for polygon, whose boundary,however, consists of $N$ edges $E_i$, and the resulting action is a sum of contributions from every edge:
\begin{equation}\label{kintermp}
S_{P}=\sum\limits_i^N \int_{E_i} d^2x  Tr(\partial_\phi \zeta \dot h_i h_i^{-1})
\end{equation}

The next step is to assemble all the above pieces of the action together and apply the condition of continuity of metric and connection across the boundary between discs and polygon.

First we convert the covariant derivative in (\ref{constsol}) into ordinary derivative by a gauge transformation
\begin{equation}\nonumber
\tilde g_i = \exp(K\phi)  g_i \ \ \ \ \ \ \tilde \xi_i = \exp(K\phi) \xi_i \exp(-K\phi)
\end{equation}
This condition violates the periodicity, so the boundary of the disk is no longer a circle but an interval.
Then disc action (\ref{kintermb}) changes to
\begin{equation}\label{kintermbt}
S_{ D_i}=\int_{\partial D_i} d^2x  Tr(\partial_\phi \tilde \xi_i \dot {\tilde g_i} \tilde g_i^{-1}),
\end{equation}
and continuity conditions for metric and connection  (\ref{constsol}) take a simple form:

\begin{equation}\label{overlap}
\tilde g_i=C_i h\Big\vert_{E_i} \ \ \ \ \ \ \tilde \xi_i = C_i \left( \zeta\Big\vert_{E_i} + \chi_i \right) C_i^{-1},
\end{equation}
where $C_i$ and $\chi_i$ are functions only of time. In the subsequent we will put $\chi_i=0$, which is possible if the positions of different particles could be related by pure rotation with no translations. This is an implementation of spherical symmetry. Substituting this into (\ref{kintermp}) and (\ref{kintermbt}), and combining them one obtains
\begin{equation}
S_{full}=S_P+\sum\limits_i^N S_{D_i}= \sum\limits_i^N \int_{E_i}Tr(\partial_\phi \zeta C_i^{-1}\dot C_i)=-
\sum\limits_i^N \int_{\partial D_i}Tr(\partial_\phi \tilde \xi_i \dot C_i C_i^{-1})
\end{equation}
The integrands are total derivatives, so the result contains contributions only from the vertices of the polygon
or endpoints of  discs boundaries.
\begin{equation}\label{sfull}
S_{full}= \sum\limits_i^N \int_{R}Tr(( \zeta_{i+1} - \zeta_{i}) C_i^{-1}\dot C_i)=-
\sum\limits_i^N \int_{R }Tr((\tilde \xi_i(2\pi)-\tilde \xi_i(0)) \dot C_i C_i^{-1})
\end{equation}
where $\zeta_i$ is the value of $\zeta$ at the i-th vertex of the polygon.

%Using overlap condition (\ref{overlap}) one can express $C_i$ as

%\begin{equation}\nonumber
%C_i=\tilde g_i  h\Big\vert_{E_i}^{-1}=\exp(K\phi)  g_i h_i^{-1}
%\end{equation}
%and the first equation in (\ref{sfull}) can be rewritten as

%\begin{equation}
%S_{full}= \sum\limits_i^N \int_{R}Tr(( \zeta_i - \zeta_{i-1})( -\dot h_i h_i^{-1} +h_ig_i^{-1}\dot g_i h_i^{-1}))
%\end{equation}

Introduce  new variables

\begin{equation}\label{nvar}
u_i=C_i^{-1}\exp(2\pi K)C_i, \ \ {\rm and} \ \ \ \ \ \ \bar \xi_i = C_i^{-1} \tilde \xi_i (0)C_i.
\end{equation}

Then taking into account that

\begin{equation}\nonumber
\tilde \xi_i (2\pi) = \exp(2\pi K)\tilde \xi_i (0)\exp(-2\pi K)
\end{equation}
and
\begin{equation}\nonumber
u_i^{-1}\dot u_i = C_i^{-1}\dot C_i - C_i^{-1}\exp(-2\pi K)\dot C_i C_i^{-1} \exp(2\pi K) C_i
\end{equation}
we can rewrite the second equation in (\ref{sfull}) as

\begin{equation}\label{sfull1}
S_{full}=
\sum\limits_i^N \int_{R }Tr(\bar \xi_i   u_i^{-1}\dot u_i )
\end{equation}

To relate $\bar \xi_i$'s for different $i$'s we use the second of the overlap conditions (\ref{overlap}) which in particular implies that
\begin{equation}\nonumber
\tilde \xi_i (0) = C_i \zeta_i  C_i^{-1}, \ \ \ \tilde \xi_i (2\pi)=\exp(2\pi K)\tilde \xi_i (0)\exp(-2\pi K)=C_i \zeta_{i+1}  C_i^{-1},
\end{equation}
from which using the definition (\ref{nvar}) one can derive
\begin{equation}
\bar \xi_{i+1}=u_i \bar \xi_i u_i^{-1},
\end{equation}
and
\begin{equation}\label{xii0}
\bar \xi_{i}=\left( \prod\limits_{j=0}^{i-1} u_j\right) \bar \xi_0 \left( \prod\limits_{j=0}^{i-1} u_j \right)^{-1},
\end{equation}
where the factors in the product are ordered from right to left.

Now introduce the holonomy around the full shell, which is the product of holonomies around every particle
\begin{equation}\label{ufull}
 U=\prod\limits_{j=0}^{N} u_i.
\end{equation}

Using an obvious identity
\begin{equation}\nonumber
U^{-1}\dot U =\sum\limits_{i=0}^N \left( \prod\limits_{j=0}^{i-1} u_j \right)^{-1} u_i^{-1}\dot u_i\left( \prod\limits_{j=0}^{i-1} u_j\right),
\end{equation}
and taking into account the relation (\ref{xii0}) we can rewrite the kinetic term of the action (\ref{sfull1}) in a simple form
\begin{equation}\label{actkinfull}
S_{full}=
 \int_{R }Tr(\bar \xi_0   U^{-1}\dot U )
\end{equation}
The action for many particle is thus collapsed to a term depending on a single variable. This became possible only because we put $\chi_i=0$ in (\ref{overlap}), making  use of spherical symmetry.

\subsection{Constraints}

To get the complete action for the shell one has to find the constraints satisfied by the variables entering (\ref{actkinfull}), and add them to the kinetic term.
In the definition of $U$
\begin{equation}\label{ueval}
 U=\prod\limits_{i=0}^{N} C_i^{-1}\exp(2\pi K_i)C_i
\end{equation}
choose a spherically symmetric anzatz for $C_i$

\begin{equation}
C_i=\exp(\bar\chi \gamma_1) \exp(\frac{2\pi i\gamma_0}{N})
\end{equation}
where $i$ (not to confuse with imaginary unit) plays the role of an angular variable, and $\bar\chi$ is an independent of $i$ boost parameter. For $K_i=M_i\gamma_0$ spherical symmetry means that $M_i=M/N$, where $M$ is the overall bare mass of the shell. Substituting this into (\ref{ueval}) one obtains

\begin{eqnarray}
 U=\prod\limits_{i=0}^{N} \exp(-\frac{2\pi i\gamma_0}{N})\exp(\frac{2\pi M}{N}(\gamma_0 \cosh \bar\chi + \gamma_2 \sinh \bar\chi) )\exp(\frac{2\pi i\gamma_0}{N}) \nonumber
 \\
 =\left( \exp(\frac{2\pi m}{N}(\gamma_0 \cosh \bar\chi + \gamma_2 \sinh \bar\chi) \exp(\frac{2\pi \gamma_0}{N})\right)^N
\end{eqnarray}
The product of two exponents in the last equation could be calculated by Campbell--Hausdorff formula $\exp(A)\exp(B)=\exp(A+B+[A,B]/2+...)$. But the terms with commutators will be of the order of $1/N^2$ and higher, and, therefore, negligible as $N \rightarrow \infty$. Finally we obtain

\begin{equation}\label{ufin}
U=\exp(2\pi ((1-M \cosh \bar\chi) \gamma_0 + M \sinh \bar\chi \gamma_2) )
\end{equation}

The conjugacy class of the above holonomy  is fixed by the boundary conditions at infinity. It can be evaluated as a Wilson loop of the angular component of the connection. The later can be expressed in terms of the ADM variables as
\begin{equation}
\omega_\phi = R'/\Lambda \gamma_0 + P_\Lambda\gamma_1
\end{equation}
and
\begin{equation}
Tr(U)=Tr(P\exp(\int d \phi \omega_\phi )) =\cos(2\pi\sqrt{(R'/\Lambda)^2-P_\Lambda^2} )=\cos(2\pi\sqrt{1-2m} ),
\end{equation}
where $m$ is the ADM mass.

On the other hand from (\ref{ufin}) it follows that

\begin{equation}\label{ufintr}
Tr(U)=\cos(2\pi\sqrt{(1-M \cosh \bar\chi)^2- (M \sinh \bar\chi)^2} )
\end{equation}

Equating two last expressions for $Tr(U)$ and solving for $M$ one finds

\begin{equation}
M=\sqrt{1+\sinh^2 \bar\chi}+\sqrt{1-2m+\sinh^2 \bar\chi}
\end{equation}

This is the Israel equation (\ref{31}) from the previous section with $\dot R /N^c = \sinh \bar\chi$.

Now we can write down the relations between canonical momenta used here and canonical momenta from the previous section:
\begin{equation}
\sinh \chi=\sqrt{1-2m}\sinh{P_R}_{out}=\sinh{P_R}_{in}
\end{equation}
and
\begin{equation}
Tr(U)=\cos(2\pi\sqrt{1-2m} )
\end{equation}
One can see that $\sinh \chi$ is always real even when $\sqrt{1-2m}$ and ${P_R}_{out}$ are complex. $U$ is also always real
and elliptic when $1-2m>0$ and hyperbolic  when $1-2m<0$. In elliptic case, $U =g^{-1}\exp(\gamma_0\phi)g$, when $\sqrt{1-2m}>0$ then $0<\phi<\pi$, and when $\sqrt{1-2m}<0$ then $\pi<\phi<2\pi$. And similar for the hyperbolic case:
$U =g^{-1}\exp(\gamma_1\chi)g$, when $i\sqrt{1-2m}>0$ then $\chi>0$, and  when $i\sqrt{1-2m}<0$ then $\chi<0$.
In other words, $U$ provides a real global parametrization  of the momentum space of the model.
This momentum space is shown on Fig.2, the four described above regions labelled as I,III,II, and VI respectively.
These regions can also be found on the Penrose diagram Fig.3, where they are labelled the same way.

\begin{figure}
\includegraphics[scale=0.70]{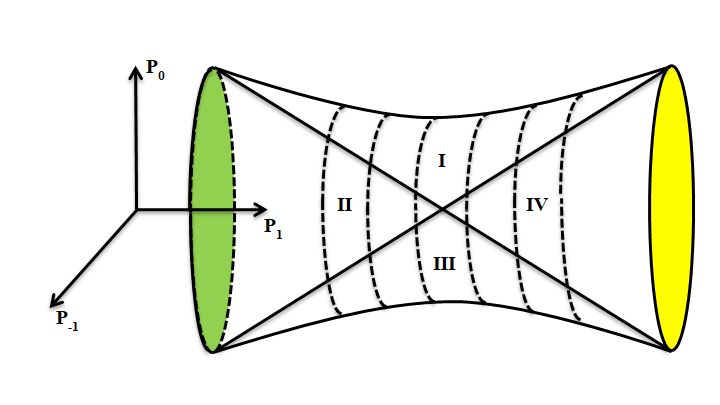}
\centering
\caption{ADS-momentum space and its four regions: Regions I and III correspond to elliptic Lorenz transformation and timelike trajectory of the shell. Regions II and IV correspond to hyperbolic Lorenz transformation and spacelike trajectory of the shell.}
\end{figure}

\begin{figure}
\includegraphics[scale=0.70]{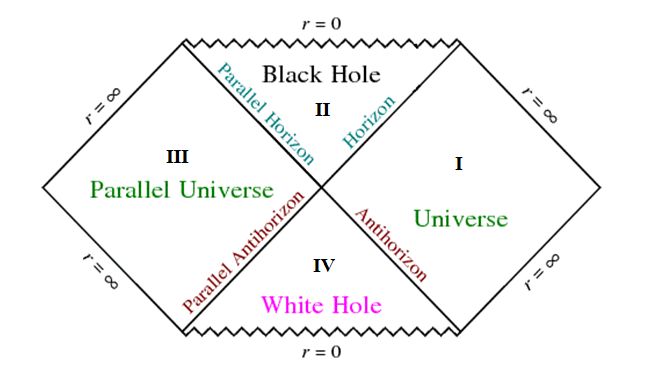}
\centering
\caption{Penrose diagram and its four regions}
\end{figure}

\subsection{Poisson brackets}

By a change of variables
the kinetic term of the action (\ref{actkinfull}) can be put in the standard canonical form

\begin{equation}\label{actkinfullc}
S_{full}=
 -\int_{R } p_a \dot q^a,
\end{equation}
where $a=0,1$ labels temporal and radial components of a coordinate (momentum).
The explicit form of this transformation is as follows:
\begin{eqnarray}
p_{-1}=Tr(U), \ \ \ \ \ p_a=Tr(\gamma_a U)
\nonumber \\
q^{a}=(p_{-1}\eta^{ab}-p_{-1}^{-1}p^a p^b)\xi_b \label{qtxi}
\end{eqnarray}

They satisfy the standard commutation relations
\begin{equation}
\{ p_a , p^b\}=0 , \ \ \{ q_a , q^b\}=0 , \ \ \{ q_a , p^b\}=\delta_a^b .
\end{equation}

It is, however, not these variables that will subsequently be used in quantization.
The reason is that because $U$ is an SL(2) group element the momenta satisfy the relation
\begin{equation}
p_{-1}^2 -p_a p^a =1.
\end{equation}
This means that the momentum space of the model is a one sheet hyperboloid or 2-dimensional anti-DeSitter space, $p_{-1}$ and $p_a$ being coordinates of 3-dimensional space in which the ADS is embedded. Momenta $p_a$ do not form a global chart on $ADS^2$. The global chart on $ADS^2$ if formed by the Euler angles which parameterize the group element $U$ as follows:
\begin{equation}\label{upar}
U=\exp(\frac{\rho}{2}\gamma_0)\exp(\chi \gamma_1)\exp(\frac{\rho}{2}\gamma_0).
\end{equation}
These are related to $p$'s as:
\begin{equation}
p_{-1}=\cos \rho \cosh \chi, \ \ \ p_0=\sin \rho \cosh \chi, \ \ \ \  p_{1}= \sinh \chi
\end{equation}
It is clear that the Euler angles together with coordinates canonically conjugate to them cannot form the standard canonical
set of variables. This is because the Euler angles parameterize a curved space which cannot be mapped onto a flat space,
and translations in such a space do not commute.

In fact the coordinates conjugate to the Euler angles are the original translation parameters $\xi^a$ from (\ref{actkinfull}). To show this we can calculate the Poisson brackets for them using the inverse of (\ref{qtxi}):
\begin{equation}
\{\xi^a, p_b \}=p_{-1}^{-1}(\delta^a_b+p^ap_b), \ \ \{\xi^a, p_{-1}^{-1} \}=p^a, \ \ \
\{\xi^a, \xi^b \}=p_{-1}(\xi^ap^b-\xi^b p^a).
\end{equation}
This implies
\begin{equation}\label{lir}
\{ U, \xi^a \}=U\gamma^a
\end{equation}
and expected non-commutativity of the coordinates.

\section{Quantization}
In this section we shall briefly describe quantization of the model using momentum (Euler angle) representation.
The exposition will basically follow \cite{mw}, but take into account that there is one less degree of freedom due to spherical symmetry, and that the constraint describing dynamics is now different.
\subsection{Quantum kinematics}
We define the kinematical states of the model as functions of $U$ from (\ref{upar})
\begin{equation}
=\Psi(U)=\Psi(\rho,\chi),
\end{equation}
single-valued functions on the entire momentum space. From the requirement of $\Psi$ being single-valued it immediately follows the periodicity property,
 \begin{equation}
\Psi(\rho+2\pi,\chi)=\Psi(\rho,\chi),
\end{equation}
which will have an important consequence on the spectra of coordinates.

Next, for defining the scalar product, we need a Lorenz-invariant measure on our momentum space. It can be inferred from the Haar measure on SL(2):
\begin{equation}
dU=\frac{1}{\pi}\sinh(2\chi)d \rho d\chi,
\end{equation}
and the scalar product is thus:
\begin{equation}
\langle\Phi, \Psi \rangle= \frac{1}{\pi}\int \sinh(2\chi)d \rho d\chi \Phi(\rho,\chi)^* \Psi(\rho,\chi).
\end{equation}

Easiest of all is to calculate the spectrum of time coordinate, $\xi^0$, which is canonically conjugate to $\rho$, and the corresponding operator is
\begin{equation}
\hat T |\rho,\chi\rangle=i\hbar\frac{\partial}{\partial\rho} |\rho,\chi\rangle
\end{equation}
its eigenstates are
\begin{equation}
|t;\psi\rangle=\frac{1}{\pi}\int \sinh(2\chi)d \rho d\chi \exp(it\rho)\psi(\chi)|\rho,\chi\rangle,
\end{equation}
where $t$ is an integer. Thus, time operator has a discrete spectrum:
\begin{equation}
\hat T|t;\psi\rangle=t\hbar  |t;\psi\rangle.
\end{equation}
Notice that it is quantized in units of the Planck length, as the above equation also contains the Newton constant, which is put to 1 in this paper.

A more interesting observable is the areal radius, $R^2=2\pi \xi_a\xi^a$, which is a lorenz-invariant quantity defining the size of the shell.  Because $\xi_a$ in (\ref{lir}) is defined as the left-invariant derivative on the group, its square is the Beltrami--Laplace operator on our momentum space:
\begin{equation}
\hat R^2 |t;\psi\rangle= 2\pi |t;\Delta \psi\rangle,
\end{equation}
where
\begin{equation}
\Delta=\hbar^2\left(\frac{1}{\sinh(2\chi)}\frac{\partial}{\partial\chi}\sinh(2\chi)\frac{\partial}{\partial\chi}
+\frac{t^2}{\cosh^2(2\chi)} \right).
\end{equation}
This operator was shown in \cite{mw} to have two series of eigenvalues. One is continuous, but separated from zero, corresponds to positive,i.e. spacelike, $R^2$
\begin{equation}
\hat R^2 |t,\lambda \rangle=2\pi(\lambda^2+1)\hbar^2 |t,\lambda \rangle,
\end{equation}
where $\lambda$ is a real number. The other is discrete, but containing zero, corresponds to negative, i.e. timelike, $R^2$
\begin{equation}
\hat R^2 |t,l \rangle=-2\pi l(l+2)\hbar^2 |t,l \rangle,
\end{equation}
where $l$ is a non-negative integer, subject to the condition $l\leq t$.

Physically the areal radius is spacelike outside an event horizon and timelike outside. In 2+1 dimensional gravity with no cosmological constant considered here there is no event horizons. So, one could say that only the continuous series is relevant to the present model.  Thinking about the entire universe being inside an event horizon is not compatible with any sensible boundary conditions at infinity.

On the other hand one could consider a more complicated model with several concentric shells which allows for spacetimes with no boundary. In such a situation  timelike areal radius is possible and the discrete series will be relevant.

However, in both cases the result for the spectrum of $R^2$ provide a regularization near $R=0$ singularity.

\subsection{Physical states}
Now we shall try to apply the hamiltonian constraint to obtain physical states. Unlike in the case of a particle, the group element repersenting the momentum (\ref{ufin} ) has a complicated dependence on the external parameter of the model, the bare mass $M$. Because of these complication we will use some implicit expressions in this section.

The analog of the Hamiltonian constraint for a particle is (\ref{ufintr}), which in terms of the Euler angles reads
\begin{equation}\label{hconst1}
\cos \rho \cosh \chi = \cos(2\pi\sqrt{1+M^2-2M\cosh\bar\chi} )\equiv p_{-1}(\bar\chi).
\end{equation}
It contains an extra parameter $\bar\chi$, which the wavefunction of the kinematical Hilbert space doesn't depend of.
To fix it one has to use another equation from the set (\ref{ufintr}), e.g.
\begin{equation}
\sin \rho \cosh \chi =\frac{(1-M)\cosh\bar\chi \sin(2\pi\sqrt{1+M^2-2M\cosh\bar\chi})}{\sqrt{1+M^2-2M\cosh\bar\chi}}\equiv p_{0}(\bar\chi)
\end{equation}
or
\begin{equation}
 \sinh \chi =\frac{M\sinh\bar\chi \sin(2\pi\sqrt{1+M^2-2M\cosh\bar\chi})}{\sqrt{1+M^2-2M\cosh\bar\chi}}\equiv p_{1}(\bar\chi)
\end{equation}
with
\begin{equation}
p_{-1}(\bar\chi)^2+p_{0}(\bar\chi)^2-p_{1}(\bar\chi)^2=1.
\end{equation}
This constraints canonically commute with each other, i.e. they are first class. So, we can start with kinematical Hilbert space as space of functions of three parameters $\Psi (\rho, \chi, \bar\chi)$ and then apply two of the above three constraints.

The solution can be written as
\begin{equation}\label{phst}
\Psi(\rho, \chi, \bar\chi)=\delta(\cos \rho \cosh \chi-p_{-1}(\bar\chi))\delta(\sin \rho \cosh \chi-p_{0}(\bar\chi))\Psi(\chi)
\end{equation}
where $\Psi(\chi)$ is an arbitrary function. As usual it is not normalizible w.r.t. the kinematical Hilbert space.

The scalar product on the physical Hilbert space can be defined in terms of functions $\Psi(\chi)$ entering (\ref{phst}) as
\begin{equation}
\langle\Psi, \Phi \rangle_{phys}=\frac{1}{\pi}\int \sinh(2\chi)d\chi d\rho d(\cosh\bar\chi)\delta(\cos \rho \cosh \chi-p_{-1}(\bar\chi))\delta(\sin \rho \cosh \chi-p_{0}(\bar\chi)) \Psi(\chi)^* \Phi(\chi).
\end{equation}
It is easy to show that in the limit $M\ll 1$ and $\chi \ll 1$ it is reduced to the standard scalar product for the states
of a relativistic particles in 1+1 dimensions,
\begin{equation}
\langle\Psi, \Phi \rangle_{phys}=\frac{1}{2\pi}\int\frac{d\chi}{\sqrt{\chi^2+M^2}}\Psi(\chi)^* \Phi(\chi),
\end{equation}
as expected.

From the above it follows, in particular, that unlike in 3+1 dimensional situation the spectrum of the energy-momentum of the
model is fully continuous. This is not surprising: in 2+1 dimensional gravity there is no Newtonian potential, so there is no potential well to hold the shell in a bounded region.

\section{Conclusion}
So far we do not have the full quantum theory of the model studied in this paper. What remains to do is to describe the dynamics, in particular calculate the transition amplitudes between different eigenvalues  of the areal radius of the shell.

What was shown is that the spectrum of areal radius in the case of timelike movement of the shell does not reach zero.
This means that  classically existing naked singularity   in quantum theory is avoided. As to singularity behind a  horizon,
which is classically attained by spacelike movement of the shell, it belongs to a discrete spectrum of the radius and, therefore, is also regularized.

The most interesting question is whether all this can be generalized for 3+1 dimensional gravity. Some results exist on quantum kinematics of a Schwarzschild black hole in a frame of a test particle \cite{drps}. It also has such features as coordinate non-commutativity and discreteness. However it cannot be generalized to a shell dynamics the way it was done in
section 3. The reason is that many body problem in 3+1 gravity is not solvable.

On the other hand almost all the work on spherical shell mentioned in the introduction was done in 3+1 spacetime dimensions.
The form of the Hamiltonian constraint (although not in global coordinates) was found. The map between phase space coordinates used there and the global ones should be analogous to that found in the end of section 3.3. The only difference is the presence of Newtonian potential. On this basis the form of the Hamiltonian constraint in global coordinates could be guessed.

\bigskip

{\bf Acknowledgements.}
We would like to thank the organizers of VI International Conference "Models in quantum field theory" dedicated to A.N.Vassiliv.

A.S.  would like to thank V.A.Berezin for extensive  conversations on his work, and D.Lyozin for collaboration on a closely related topic.

This work was partially supported by
RFBR projects 16-02-00348 and 18-02-00264 (A.A.)

%\bibliographystyle{../my2beznazv}
%\bibliography{../paston-grav-r,../paston-grav-e}

\end{document}